%% file: main.tex
\documentclass[sigconf]{acmart}
\usepackage{amsmath,amssymb,amsfonts}
\usepackage[ruled,vlined,linesnumbered]{algorithm2e}
\usepackage{algpseudocode}
\usepackage{array}
\usepackage{booktabs}
\usepackage{comment}
\usepackage{enumitem}
\usepackage{graphicx}
\usepackage{listings}
\usepackage{multirow}
\usepackage{pifont}
\usepackage{subfigure}
\usepackage{seqsplit}
\usepackage{textcomp}
\usepackage{tikz}
\usepackage{url}
\usepackage{xcolor}
\usepackage{stfloats}

\AtBeginDocument{%
  \providecommand\BibTeX{{%
    \normalfont B\kern-0.5em{\scshape i\kern-0.25em b}\kern-0.8em\TeX}}}

\copyrightyear{2021}
\setcopyright{iw3c2w3}
\acmYear{2021}
\acmConference[WWW '21]{Proceedings of the Web Conference 2021}{April 19--23, 2021}{Ljubljana, Slovenia}
\acmBooktitle{Proceedings of the Web Conference 2021 (WWW '21), April 19--23, 2021, Ljubljana, Slovenia}
\acmPrice{}
\acmDOI{10.1145/3442381.3450007}
\acmISBN{978-1-4503-8312-7/21/04}

\setenumerate[1]{itemsep=0pt,partopsep=0pt,parsep=\parskip,topsep=1.5pt}
\setitemize[1]{itemsep=0pt,partopsep=0pt,parsep=\parskip,topsep=0pt}
\setdescription{itemsep=0pt,partopsep=0pt,parsep=\parskip,topsep=1.5pt}

\definecolor{codegreen}{rgb}{0,0.6,0}
\definecolor{codegray}{rgb}{0.5,0.5,0.5}
\definecolor{codepurple}{rgb}{0.58,0,0.82}
\definecolor{backcolour}{rgb}{0.95,0.95,0.95}
\definecolor{keywordsColor}{rgb}{0.000000, 0.000000, 0.635294}

\lstdefinestyle{mystyle}{
    basicstyle=\scriptsize\ttfamily,
    aboveskip=0.3ex,
    belowskip=0.25ex,
    breakatwhitespace=false,         
    breaklines=true,      
    columns=fullflexible,           
    captionpos=b,                    
    keepspaces=true,                                     
    numbersep=4pt,                  
    showspaces=false,                
    showstringspaces=false,
    showtabs=false,
    tabsize=2,
    language=XML,
    otherkeywords = {typeattributeset,and,not,neverallow,allow,never,typeattribute,\$},
    keywordstyle=\color{keywordsColor}\bfseries,
    backgroundcolor=\color{backcolour}, 
    commentstyle=\color{codegreen}
}

\newcommand{\don}[1]{\textcolor{green}{Don:#1}}

\newcommand{\tool}{SEPAL\xspace}
\newcommand{\ie}{\textit{i.e.}\xspace}
\newcommand{\eg}{\textit{e.g.}\xspace}
\newcommand{\etal}{\textit{et al.}\xspace}

\newcommand{\textsfs}[1]{\textsf{\small #1}}

\newcounter{summary}

\begin{document}

\date{}
\title{SEPAL: Towards a Large-scale Analysis of SEAndroid Policy Customization}

\author{Dongsong Yu$^{1,2}$, Guangliang Yang$^{3}$, Guozhu Meng$^{1,2*}$, Xiaorui Gong$^{1,2*}$, Xiu Zhang$^{1,2}$, Xiaobo Xiang$^{1,2}$, Xiaoyu Wang$^{1,2}$, Yue Jiang$^{1,2}$, Kai Chen$^{1,2}$, Wei Zou$^{1,2}$, Wenke Lee$^{3}$ and Wenchang Shi$^{4}$ 
}
\affiliation{%
\institution{$^{1}$Institute of Information Engineering, Chinese Academy of Sciences, China\\
$^{2}$School of Cyber Security, University of Chinese Academy of Sciences, China\\
$^{3}$Georgia Tech, USA\\
$^{4}$Renmin University of China, China\\
}
}
\email{{yudongsong,mengguozhu,gongxiaorui,zhangxiu,xiangxiaobo,wangxiaoyu,jiangyue,chenkai,zouwei}@iie.ac.cn}
\email{guangliang.yang11@gmail.com,wenke.lee@gmail.com,wenchang@ruc.edu.cn}
\renewcommand{\shortauthors}{D. Yu, G. Yang, G. Meng, X. Gong, X. Zhang, X. Xiang, X. Wang, Y. Jiang, K. Chen, W. Zou, W. Lee, and W. Shi}

\def\@copyrightpermission{\relax}
\input{1-abstract}

\maketitle

\newcommand\blfootnote[1]{%
\begingroup 
\renewcommand\thefootnote{}\footnote{#1}%
\addtocounter{footnote}{-1}%
\endgroup 
}
\blfootnote{* Guozhu Meng and Xiaorui Gong are the corresponding authors.}

\input{2-introduction_new}

\input{3-background}
\input{problem}

\input{4-methodology_new}
\input{5-evaluation}
\input{6-analysis}
\input{6-discussion}
\input{7-relatedwork}

\input{8-conclution}

\input{Acknowledgement}

\bibliographystyle{ACM-Reference-Format}
\bibliography{reference}

\end{document}

%% file: 1-abstract.tex
\begin{abstract}
Nowadays, SEAndroid has been widely deployed in Android devices to enforce security policies and provide flexible mandatory access control (MAC), for the purpose of narrowing down attack surfaces and restricting risky operations.
Generally, the original SEAndroid security policy rules are carefully and strictly written and maintained by the Android community. However, in practice, mobile device manufacturers usually have to customize these policy rules and add their own new rules to satisfy their functionality extensions, which  breaks the integrity of SEAndroid and causes serious security issues. 
Still, up to now, it is a challenging task to identify these security issues due to the large and ever-increasing number of policy rules, as well as the complexity of policy semantics.

To investigate the status quo of SEAndroid policy customization, we propose \tool, a universal tool to automatically retrieve and examine the customized policy rules. 
\tool applies the NLP technique and employs and trains a wide\&deep model to quickly and precisely predict whether one rule is unregulated or not.
Our evaluation shows \tool is effective, practical and scalable. We verify \tool outperforms the state of the art approach (i.e., \textsc{EASEAndroid}) by 15\% accuracy rate on average. In our experiments, \tool successfully identifies 7,111 unregulated policy rules with a low false positive rate from 595,236 customized rules (extracted from 774 Android firmware images of 72 manufacturers). We further discover the policy customization problem is getting worse in newer Android versions (e.g., around 8\% for Android 7 and nearly 20\% for Android 9), even though more and more efforts are made. Then, we conduct a deep study and discuss why the unregulated rules are introduced and how they can compromise user devices. 
Last, we report some unregulated rules to seven vendors and so far four of them confirm our findings.
\end{abstract}

%% file: 2-introduction_new.tex
\section{Introduction}
\label{section:Introduction}
Nowadays, SEAndroid (Security-Enhanced Android) has been widely applied in Android devices to shrink attack surface. It is fully enforced since Android 5, and now serving and protecting more than 98\% Android devices as of Sep, 2020~\cite{androidmark}. SEAndroid is derived from SELinux (an effective security enforcement module in the Linux kernel)~\cite{smalley2013security}, and it provides robust and flexible mandatory access control (MAC) on sensitive resources and operations. 
The effectiveness of SEAndroid is often highly determined by the correctness and completeness of the security policy, a set of manually defined entries named \emph{rules}. With well-developed policy rules, SEAndroid can easily and conveniently confine system services, control sensitive data access, reduce the effects of malicious apps, and protect users from potential flaws in apps. For this reason, in the official Android source code, \ie, Android Open Source Project (AOSP), a solid base policy is carefully defined and well maintained. This base policy provides effective protection on apps and system services, which is reflected and verified in a recent study~\cite{us-17-blackhat}:  almost 50\% of the kernel bugs can be blocked by the official policy in Android 8.

However, in practice, Android devices in the wild are usually customized by manufacturers. To ensure that the customized features can work properly, they need to add device-specific rules into their devices. Unfortunately, the customized policy may undermine the original defense provided by SEAndroid.
For example, origin policy strictly confines the access of unprivileged processes to sensitive device nodes, thus most of the kernel driver vulnerability cannot be triggered by an adversary to escalate the privilege.
Yet a customized rule in some devices breaks the confinement. For example, it allows some unnecessary processes to access the vulnerable driver of a MediaTek component (known as CVE-2020-0069)~\cite{cve-2020-0069}, which can lead to an attack of local privilege escalation. 
For convenience, we refer to the potentially risky or unnecessary rules as \emph{unregulated rules} in this study.

To understand the status quo in policy customization, we need a general method to identify the unregulated rules in the wild. However, it is also a challenging task. 
First, the Android fragmentation problem greatly hinders policy analysis. The representation of policy may be different across devices, causing that the rules of the same semantics look different (see Section \ref{subsection:policy diversity}). 
Second, it is difficult to obtain sufficient semantics to evaluate the rationality of the target rules merely based on the static representation of the rules.
Third, there is no clear ground truth to judge whether a rule is permissive or not. The determination heavily relies on the expertise and the case-by-case analysis sometimes can be subjective. Even though we can write a script to query the unregulated rules based on the patterns summarized by prior studies~\cite{reshetova2015characterizing,Chen:2017:ASP:3134600.3134638}, it cannot find unregulated ones of previously unknown categories. In fact, since the policy has evolved a lot, the old patterns hardly appear in the latest versions of Android.

To meet these challenges, in this paper, we present a novel unregulated policy detection solution, termed as \textbf{SE}Android \textbf{P}olicy \textbf{A}na\textbf{l}yzer (\tool), to automatically vet the massive number of customized rules.
The rationale of this study is to train an explicit boundary between the unregulated and ordinary ones based on the solid policy rules defined in AOSP via a learning-based approach. We aim to highlight the outliers in the dataset from the perspective of statistics and then manage to explain why they are classified as unregulated thus to identify previously unknown patterns of the unregulated rules.
In particular, \tool proceeds in three phases. 
First, given a firmware image, \tool retrieves all the rules regardless of policy formats and versions. To address the fragmentation problem and unify the expressions of rules, \tool transforms the rules defined in different representations into a uniform format named \textit{atomic rules}. 
To enrich the semantics in atomic rules, \tool then generates informative features to represent the atomic rules from the policy attributes and user IDs of the corresponding subjects. We further manage to extract some semantic features generated by NLP (Natural Language Processing) from official policy comments to collect domain-specific knowledge about the corresponding processes. 
Furthermore, we use the atomic rules collected from AOSP to model the complicated correlations among the subjects and objects in SEAndroid via machine learning. 
To achieve high accuracy, we enrich the original imbalanced dataset and use a compound model jointed with a linear model and a deep neural network (DNN) to learn the feature combinations in the official policy, which is inspired by modern recommendation systems.
It yields a classifier to help us highlight the unregulated rules in the huge amount of the customized rules.
We also design a baseline algorithm based on the traditional machine learning methods for comparison, which are used in prior research on SEAndroid. The results show that \tool performs more accurately in our task.

To evaluate the performance of \tool, we collect 595,236 customized rules from 774 Android images of 72 manufacturers. 
\tool retrieves nearly 3.5 million of atomic rules from these images. By applying \tool on these rules, we find that more than 12\% of customized atomic rules are unregulated, which comes from 7,111 unique policy rules defined by manufacturers. In our analysis, we find that things improved a lot in the Android 7 era because Google paid their efforts to break down some coarse types and dismissed the usage of some error-prone attributes proposed in earlier research~\cite{reshetova2015characterizing,Chen:2017:ASP:3134600.3134638}. However, things seem to be going from bad to worse nowadays - the percentage has risen to 17.76\% in Android 9. 
We find that even well-known manufacturers such as Samsung and Huawei struggle to keep up with the evolution of official policy, not to mention other manufacturers of less scale. 
Based on the results presented by \tool, we further conduct an in-depth analysis of these unregulated rules. 
We summarize four previously unknown reasons why these rules are introduced such as the misuse of attributes, the debugging related rules, deprecated rules, and the excessive permissions granted to untrusted domains.
Furthermore, to demonstrate how the rules can downgrade the defense of official SEAndroid, we show the security impacts of these unregulated rules, including the exposure of system services and sensitive device nodes, extending the capabilities of malicious apps, and providing extra paths for vulnerability exploitation. 
To illustrate the security impact, we also conduct attacks for a case study in which: \textit{1)} we control the screen lock password of the device, \textit{2)} communicate with the camera without permission request, which is acknowledged by corresponding vendors. 
We have contacted seven vendors about the issues and four of them confirm our findings. 
To the best of our knowledge, this is the first work to perform such a large-scale measurement on policy customization. 
We hope our findings may help manufacturers improve their policy writing. 

The contributions of this paper are concluded as follows:
\begin{itemize}[leftmargin=*]
\item\textbf{New techniques.} We propose a universal methodology, \tool, to overcome the challenges in automatic analysis of massive customized rules. \tool represents the rules in a fine-grained format and utilizes NLP techniques and deep learning to effectively detect unregulated rules. It helps a comprehensive and scalable discovery of security issues in policy customization.
\item\textbf{New findings.} \tool identifies 7,111 unique unregulated rules from 3.5 million atomic rules collected from 70 manufacturers. We then perform a large-scale measurement on policy customization across versions and vendors. We shed light on a number of important insights including why these rules are unregulated and what security impacts they can cause.
\end{itemize}

%% file: 3-background.tex
\section{Background}
\label{section:Background}
\subsection{Overview of SEAndroid}
\label{subsection:SEAndroid}
\noindent \textbf{SEAndroid Concepts. }
To enhance security by implementing \emph{mandatory access control (MAC)}, Smalley \etal customized SELinux for Android, a.k.a. SEAndroid~\cite{smalley2013security}, which is ported by Google in Android 4.3 and has fully been enforced since Android 5. 
Traditional MAC contains multiple security mechanisms: role-based access control (RBAC)~\cite{sandhu1998role}, Multi-Level Security (MLS)~\cite{mccullough1987specifications} and Type Enforcement (TE)~\cite{badger1995practical}.
Therefore, all the subjects (e.g., processes) and objects (e.g., files, sockets) involved in SEAndroid inherit security labels from these mechanisms, in the format of  ``\texttt{\seqsplit{user:role:type:security level}}''. The fields \texttt{user} and \texttt{role} are mainly used for RBAC and seldom used in SEAndroid, nor the field \texttt{security level} used for MLS. The access control of SEAndroid is mainly achieved by TE, which uses \texttt{type} field to confine the behaviors in a system via a large number of type enforcement rules.

In general, a type enforcement rule (abbreviated as rule) conforms to  ``\textit{allow domain type : class permission}'', which defines the allowable actions of a subject in the system. 
In particular, the \emph{domain} field specifies the label of the subjects (e.g., processes), \emph{type} indicates the labels of objects (e.g., files and sockets), \emph{class} is the class of objects, and \emph{permission} presents specific operations performed on the objects. 
For instance, a rule ``allow untrusted\_app app\_data\_file: file \{open read\};''
allows all the processes labeled by ``untrusted\_app''  to \emph{open} and \emph{read} ordinary \emph{files} labeled by ``app\_data\_file''. All the rules are written by system developers and saved in \texttt{*.te} files in the source code tree. 

In order to write and manage those rules correctly and efficiently, most types defined in SEAndroid have several attributes. 
For instance, \emph{untrusted\_app} has multiple attributes including \emph{domain}, \emph{appdomain}, and \emph{netdomain}, indicating that the untrusted app is a process, an app process, and a process with network access, respectively. The rules defined by attributes are valid for all the subjects or objects that have this attribute. 
Policy writers can also use macros to provide reusable pieces of the statement. For example, \emph{rw\_file\_perms} is a macro standing for a list of permissions - \emph{{getattr open read ioctl lock map append write}}, \emph{binder\_use(domain)} can generate all the rules related to the usage of Binder IPC in a single line for the target domain.   
Moreover, there are two types of imperative rules in the source code tree of SEAndroid--\emph{allow rules} and \emph{neverallow  rules}. A \emph{neverallow rule} will not be compiled into devices, but it will interrupt compilation when there exist \emph{allow rules} violating the neverallow entries. These rules are mainly used to avoid the introduction of unregulated rules. 

These allow rules in \texttt{*.te} files are finally compiled into a firmware image and initialized when the system boots up.  
All the access attempts that are not explicitly allowed in the policy will be denied by SEAndroid, even if the process is running as root:
\begin{lstlisting}
$ id
uid=0(root) gid=0(root) groups= ... context=u:r:untrusted_app:s0:c512,c768
$ ls /
ls: /: Permission denied
$ logcat |grep avc
11-06 06:41:49.193  2810  2810 W ls:type=1400 audit(0.0:19): avc:denied {read}
for name="/" dev="sda43" ino=2 scontext=u:r:untrusted_app:s0:c512,c768 
tcontext=u:object_r:rootfs:s0 tclass=dir permissive=0
\end{lstlisting}

\subsection{Policy Diversity}
\label{subsection:policy diversity}
The policy rules may differ a lot from the following perspectives.

\noindent \textbf{Version Diversity. }
Both SELinux version and SEAndroid version are used for indicating the status of the policy.
From the perspective of the SELinux version, it changes when new syntax features are added to the policy. 
It may lead to the changes of policy syntax as well as file structure in the binary policy. For example, Android 5 uses the SELinux version of 26.0, while Android 6 uses 30.0, which leads to the changes of some macros and reserved words.

From the perspective of the SEAndroid version, it is synchronized with the version of Android. It indicates the changes in the policy with the evolution of the operating system. For instance, in the SEAndroid policy rules before version 27, all processes can access the ashmem device by directly opening the character device: ``\textsfs{allow domain ashmem\_device: chr\_file rw\_file\_perms}.''
However, on Android 9 and SEAndroid 28, processes have to use a native API for the same purpose, and such rule is modified to 
``\textsfs{allow domain ashmem\_device: chr\_file getattr read ioctl lock append write }.''
Thus opening ashmem device in \texttt{/dev} from a normal user-level process will be denied in Android 9. 
In fact, the official policy usually changes a lot when the Android version evolves. 

\noindent \textbf{Policy Format Diversity. }
Policies with different file formats vary in presentation. 
The following code snippets in different policy files show how one rule is expressed in different styles.
\begin{lstlisting}
#TE style:
allow { appdomain -isolated_app } app_data_file :file r_file_perms
#CIL style:
allow base_typeattr_97 app_data_file (file (getattr open read ioctl lock map ))
#Binary style:
allow bluetooth app_data_file: file getattr open read ioctl lock map ;
allow platform_app app_data_file: getattr open read ioctl lock map;
allow untrusted_app app_data_file: file getattr open read ioctl lock map; ...
\end{lstlisting}

Rules in the AOSP are saved in \texttt{*.te} files classified by domains. They are mostly written in macros, attributes, and logical symbols (such as ``*'' for all, ``-'' for exception) for the convenience of policy writing.
While the binary policy built into devices could be decompiled by \textit{setools}~\cite{setools}. 
The corresponding rules defined by attributes in source code could be expressed as multiple entries in the decompilation result of \textit{setools}. It means that the granularity of rules may differ between binary policy and source code policy. In addition, \emph{neverallow rules} will not be compiled into the binary policy.

With the introduction of Project Treble~\cite{projecttreble} in Android 8.0, a new format of policy named Common Intermediate Language (CIL) is used as the intermediate representation betweeen \texttt{*.te} policy files and binary policy files -  
in addition to a compiled binary policy, the devices also contain multiple compiled CIL files.
Rules defined in CIL files maintain the order and the shape of both \emph{allow} and \emph{neverallow} rules defined in source code (\texttt{*.te} files). Thus CIL files contain much more information than that obtained from binary policy using \textit{setools}.  For instance, the above mentioned attribute \emph{base\_typeattr\_97} is assigned to all the app-level types except \emph{isolated\_app}, which retains the shape of the negation expressions.

\noindent \textbf{Policy Polysemy. }
Rules with identical syntax in \texttt{*.te} or binary policy files might have different meanings in different devices due to the customization of attribute definitions. For example, the definition of attribute \emph{hal\_audio} may differ between the Pixel device and some third-party devices, resulting in that the rule below may have a different meaning:``\textsfs{allow hal\_audio audio\_device:chr\_file ioctl}''
Such a rule is reasonable in AOSP, but in some third-party devices, the attribute \emph{hal\_audio} is additionally assigned to domain \emph{hal\_ir\_default} and domain \emph{hal\_irsi\_default}. The two domains are mostly assigned to the infrared-related server process. It will grant more permissions than the original rule (shown in Section~\ref{subsection:unregulated reasons}).

%% file: problem.tex
\begin{figure}
  \centering
  \includegraphics[width=0.38\textwidth]{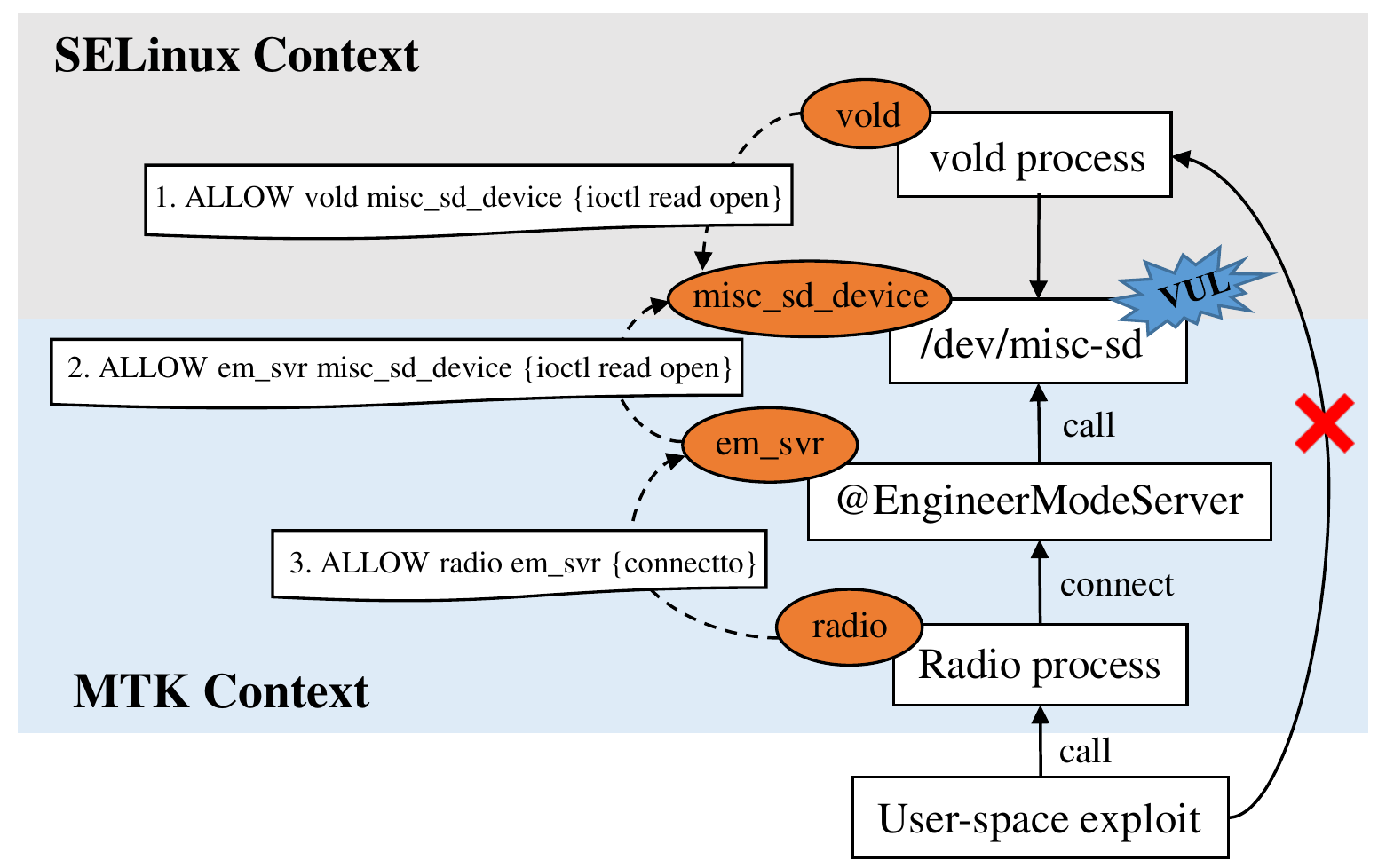}
  \vspace{-0.2cm}
  \caption{Example of security risks amid customization}\label{fig:motivation}
  \vspace{-0.2in}
\end{figure}

\section{Problem Statement}\label{sec:problem}
In this section, we present the problem to solve in this paper.
The customization of policy may bring severe security risks and destructive effects on the SEAndroid-enforced devices. 
Figure ~\ref{fig:motivation} shows an illustrative example that one vulnerability can be exploited due to an improper addition to the original policy. 
Specifically, an out-of-bounds access vulnerability (CVE-2015-6637) has been found in the driver of MediaTek (MTK) storage. The driver exposes a device node \textsfs{/dev/misc-sd} labeled by \emph{misc\_sd\_device} and regulated by original policy where only the \emph{vold} domain can access the vulnerable process (Rule 1). 
As a result, unprivileged processes are unable to exploit the vulnerability in this way when SEAndroid is enforced. 
However, several MTK devices write two customized rules (Rule 2 and 3) which enables another path to reach the vulnerability point. 
As such, user-space exploits can interact with process ``radio'', connect to the \emph{em\_svr} daemon and subsequently trigger the vulnerability in kernel. 
Attributed to compatibility issues and insufficient policy validation, vendors are prone to write these unnecessary or risky rules that enlarge the attack surface, which can be further exploited by malware~\cite{issta2016smart,tifs2018spread}.

To simply yet precisely state this problem, we provide a formal definition for these policy rules and then discuss the violation.

\noindent\textbf{Definition 1 (SEAndroid Policy Rule)} The policy rule in Android can be represented as a six tuple $(Op, \mathcal{D}, \mathcal{T}, \mathcal{C}, \mathcal{P}, Attr)$, where $Op$ indicates allow or neverallow, $\mathcal{D}$ is the set of domains, $\mathcal{T}$ is the set of component type, $\mathcal{P}$ presents all pre-defined permissions, and $Attr$ is a subset of domains $\mathcal{D}$ or types $\mathcal{T}$. As such, $Attr~\subset~\mathcal{D} \cup\mathcal{T}$.  

Intuitively, a rule can be represented as $r = \langle op, d|attr, t|attr, c, \{p_i\} \rangle$, where $op \in Op$, $d \in \mathcal{D}$, $t \in \mathcal{T}$, $c \in \mathcal{C}$, $p_i \in \mathcal{P}$. Additionally, $attr$ refers to a set of domains or types. 
Although $attr$ can ease policy writing, it raises the difficulty of policy validation and analysis due to policy diversity (Section~\ref{subsection:policy diversity}).
To this end, we propose ``atomic rule'' as the new metric for representing policy rules. The rationale here is that we noticed the definitions of attributes may vary across devices but the specified types remain stable.
Specifically, atomic rule can be defined as:

\noindent \textbf{Definition 2 (Atomic Rule)} It is a concrete instance of policy rule and can be represented as $r_a = \langle d, t, c, p \rangle$. Let $\phi$ be a permission identifier where $\phi(r_a) \in \{allow, neverallow\}$. This rule specifies only one permitted or forbidden behavior of a certain domain toward a single target. So it is irreducible, that is, cannot be further decomposed into finer-grained rules.

To identify the security risks brought by manufacturers' customization and understand how they are introduced, we intend to learn the boundary of permitted and forbidden correlations of the types defined in the AOSP policy and determine the violations between customized rules and these correlations.  
Formally put, the study aims to generate a discriminator $\mathcal{DC}$ that outputs the allow or neverallow permission given by an atomic rule, \ie, $\mathcal{DC}(r_a) \in \{allow, neverallow\}$. 
As a consequence, one violation (\ie, \textit{unregulated rule}) is found if $\mathcal{DC}(r_a) \neq \phi(r_a)$. 
Further, the violations highlighted by \tool can help us not only identify the previously unknown patterns but better understand how to get rid of the introduction of the unregulated rules.

\begin{figure*}[!t]
	\centering
	\includegraphics[width=0.75\linewidth]{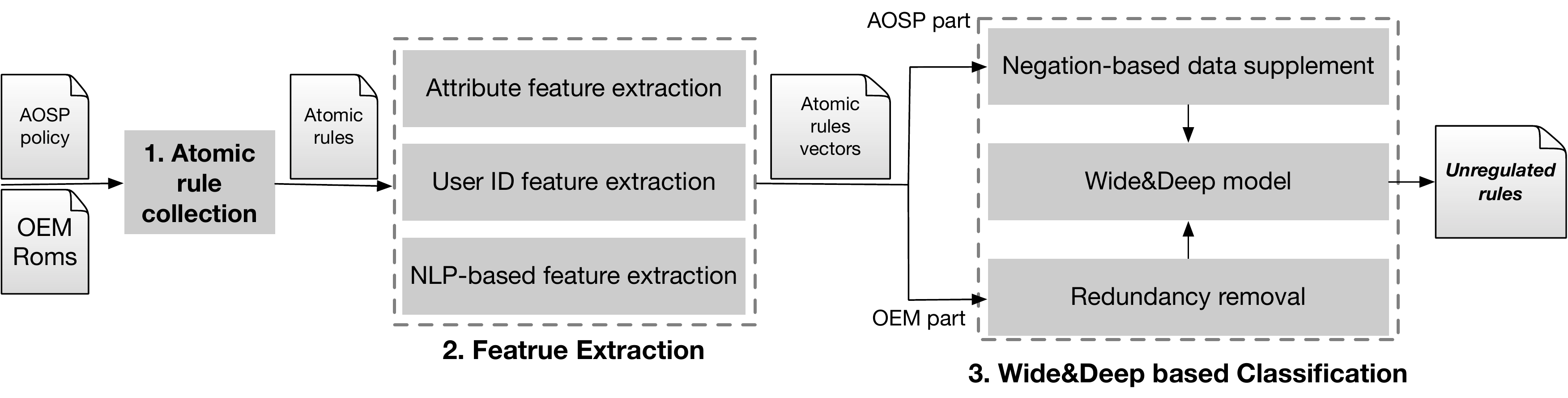}
	\vspace{-0.3cm}
	\caption{\tool's workflow}
	\label{fig:workflow}
	\vspace{-0.4cm}
\end{figure*}

%% file: 4-methodology_new.tex
\section{Methodology}
\label{section:Methodology}
Figure~\ref{fig:workflow} shows the workflow of \tool, and we describe how the three components, \ie, Atomic Rule Collection, Feature Extraction, and Wide\&Deep Based Classification, work together to identify the unregulated rules in manufacturer images in this section.

\subsection{Atomic Rule Collection}
Considering the diversity of policy file formats and policy granularity, at first we express the rules in different file formats from AOSP and firmware images into the atomic form. We develop a universal parser to obtain the metadata, \ie, the six tuples $(Op, \mathcal{D}, \mathcal{T}, \mathcal{C}, \mathcal{P}, Attr)$, from the policy files in AOSP and firmware images and then translate the metadata into atomic rules.
For policy rules in the manufacturer images from different versions, we only need the \emph{allow rules}. Thus we can directly extract the original policy rules from the binary policy files or CIL files.
Existing tools such as~\cite{sdat2img,splitupdate} can be used for firmware extraction according to how these images are packed. 
The policy rules after Android 8 are saved in CIL files and located in system and vendor partition separately.   
It is trivial to extract the tuples after merging these CIL files.
On the earlier versions of Android, the binary policy files are packed in the ramdisk of the boot image. 
The parser uses \textit{setools} to decompile the binary policy and retrieves \emph{allow rules} and attributes in the binary policy.

We also need AOSP rules for model training. It extra requires the \emph{neverallow rules} as the negative data record, thus we cannot use binary policies because the \emph{neverallow rules} are never compiled into binary. Accordingly, directly parsing \texttt{*.te}  files in AOSP is cumbersome because they are written in complicated macros and logical symbols, and the syntax of these rules may also evolve across versions. 
As such, we use CIL files for constructing the training dataset. Moreover, the CIL files nearly maintain all the semantics defined in \texttt{*.te}, thus could be used to enrich our initial dataset (shown in Section \ref{subsection:Wide Deep}).
But the CIL files are only available after Android 8. For former versions, we modified the \textsc{checkpolicy}~\cite{checkpolicy}, an open-source \texttt{te} compiler, to add compatibility support to compile the CIL files from earlier AOSP source code.
Then the metadata could be easily retrieved from the CIL policy files. 

After obtaining the tuples from AOSP and firmware images, we recursively expand all the subjects and objects defined by these attributes until all the attributes are replaced in the rules. 
Then we split the permissions separately to obtain atomic rules. 
\vspace{-2mm}
\subsection{Feature Extraction}
\label{subsection:Feature Extraction}
To better represent the semantics between elements in an atomic rule, we distill three types of features from these rules. 

\noindent\textbf{F1 Basic semantics in atomic rules}.
Every subject, object, class, and permission that has appeared in the policy will be directly mapped to the vector space after one-hot encoding.
Furthermore, some attributes can intuitively characterize the types in the rules but they are dismissed during the process of decomposing original rules into atomic rules. 
Specifically, the following attributes are represented as six boolean features: \textit{domain} indicates whether the subject is a process, \textit{MLS} indicates whether the process can override Multi-Level Security~\cite{sandhu1993lattice-based} restrictions, \textit{core} indicates whether the type is defined by the platform, \textit{app} indicates whether the subject is an app process, \textit{net} indicates whether the subject can access network, and \textit{untrusted} indicates whether the untrusted code can be executed in the process context. 

\noindent\textbf{F2 User ID of running processes.}
The \textit{user ID (UID)} indicates the privilege level of a process. 
Obtaining the UID from a running device is trivial. Prior research~\cite{Chen:2017:ASP:3134600.3134638} directly uses ``\textsfs{adb shell ps -Z}'' to collect this information via an adb connection from the rooted devices, which is not feasible in large-scale analysis.
To this end, we develop a static approach to associate user IDs and types in policy without a runtime environment.
The UID of app-related subjects is trivial to obtain - a file named \emph{seapp\_context} indicates all the user accounts of these subjects.
For system daemons, inferring their UIDs requires the correlated information saved in \emph{file\_context} in the system, init script files \emph{*.rc}, and \textsfs{typetransition} entries in the policy.
Take the type \emph{mediadrmserver} as an example. To obtain its user account, at first we look up the policy files and obtain an entry ``\textsfs{typetransition init mediadrmserver\_exec process mediadrmserver}'', which shows the type \emph{init} will be converted into \emph{mediadrmserver} when it executes files labeled by \emph{mediadrmserver\_exec}. Then \emph{file\_context} shows the type \emph{mediadrmserver\_exec} is only assigned to \textsfs{/system/bin/mediadrmserver}. Finally, the following entry is spotted in \emph{mediadrmserver.rc}:
\begin{lstlisting}
service mediadrm /system/bin/mediadrmserver ... user media ...
\end{lstlisting}
It concludes that the \emph{mediadrm} process labeled as \emph{mediadrmserver} runs as \emph{media}.

\noindent\textbf{F3 Rule comments. }
Sometimes similar operations may be expressed with different \emph{classes} or \emph{permissions}. Take the rules defined in file \emph{hal\_wifi.te} and \emph{app.te} for example. 
\begin{lstlisting}
========= hal_wifi.te ========= 
# Allow hal_wifi to send dump information to dumpstate:
allow hal_wifi dumpstate:fifo_file write;
========= app.te ========= 
# Allow apps to send dump information to dumpstate:
allow appdomain dumpstate:fd use;
allow appdomain dumpstate:unix_stream_socket {read write getopt getattr};
\end{lstlisting} 
From the comments, both the \emph{app} and \emph{hal\_wifi} can send dump information to \emph{dumpstate}, even though they use different IPC channels. 
To characterize these latent correlations between subjects, we need to parse and vectorize the comment text for the corresponding types.
However, the comments are usually written arbitrarily, some of them even have little to do with the semantics of the rules, such as bug IDs or corresponding shell commands.
Considering that the comment we concern about essentially defines the actions of a subject and the resources it can access, we manage to extract the keywords about \emph{who-does-what} from the comments and represent them into vectors via NLP techniques.

\begin{figure}[t]
	\centering
	\includegraphics[width=0.4\textwidth]{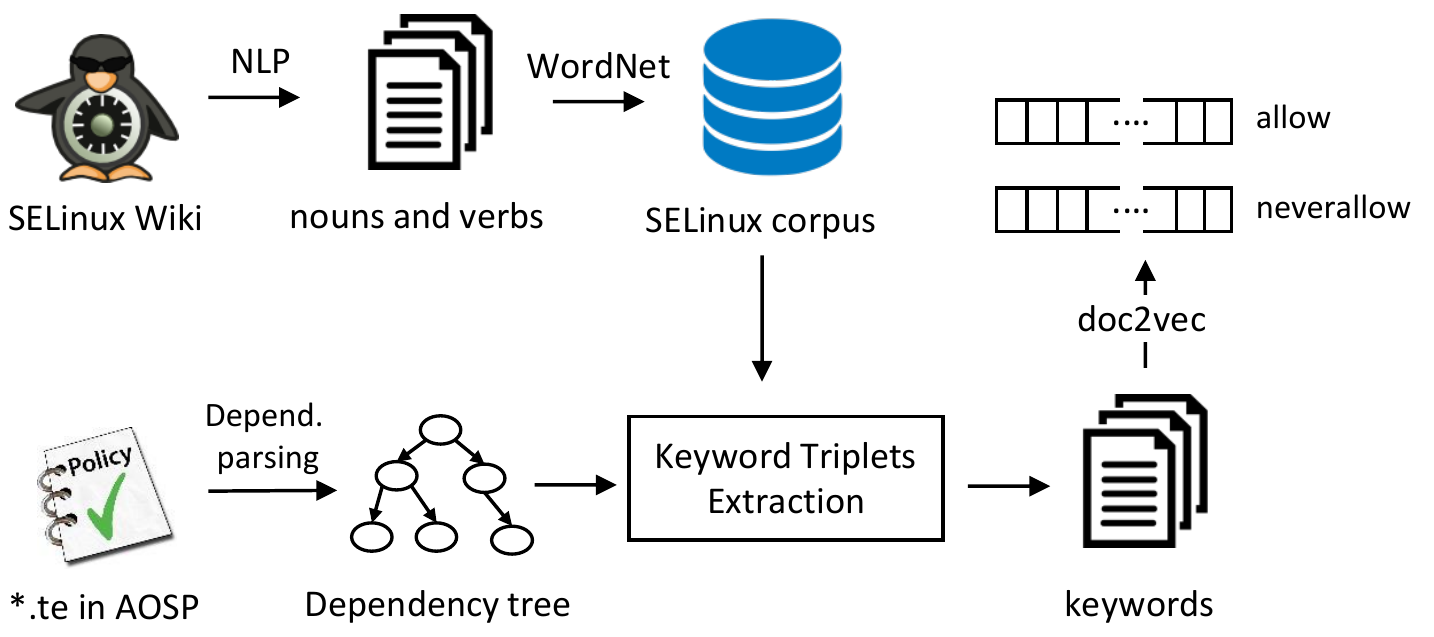}
	\vspace{-0.2in}
	\caption{The workflow of rule comment vectorization}
	\label{fig:NLP workflow}
\end{figure}

Figure~\ref{fig:NLP workflow} shows how we represent the comments of a subject into two vectors (\ie, \emph{allow} and \emph{neverallow} vectors). 
First, we collect a policy-related corpus from the SELinux Wiki page~\cite{selinux}. Based on the definitions of object classes and permissions used by SEAndroid, we extract all the verbs (\eg, {open}, {read}) and general resources (\eg, file, sockets) related to access events. Then we use \textsc{WordNet}~\cite{miller1998wordnet} to obtain the synonyms of these verbs to enrich the corpus. 
Last, we obtain 515 verbs and 49 resources, and this corpus can help to identify the sentences related to the access events and filter the irrelevant words in the comments. But simply using a keyword search method is prone to produce false positives due to the polysemy in English. 
So we perform the dependency parsing to get the words' dependency tree with the part of speech (POS) tags. Then, we implement a keyword triplets extractor based on the principle of finding \emph{SVOs} in NLP, which takes the dependency tree mentioned above as inputs and extracts \emph{subject–verb–objects} from a sentence. 

\begin{algorithm}[t]
\setlength{\textfloatsep}{0pt}
    \small
	\caption{Find keyword triplets in a sentence}\label{algorithm:FindKws}
	\KwIn{$dt$: the dependency tree of the sentence}
	\KwOut{$kt$: the keyword triplets}
	$Verbs,Objects \gets \Call{getVO}{dt}$; \\
	\For {v $\in$ Verbs}{ 
		\If {\Call{inActionCorpus}{v}}{
			$act \gets v $; $resources \gets \Call{getObjs}{v}$; \\
			\For {res $\in$ resources}{
				$comp \gets \Call{getComp}{res}$; $tr = \{act, res, comp\}$; \\
				\If {$tr \notin ~kt$}{
					$kt \gets kt ~\cup~tr$; \\
				}
			}
		}
	}
	\For {obj $\in$ Objects}{
		\If {\Call{inResourceCorpus}{obj}}{
			$res \gets obj $; $comp \gets \Call{getComp}{obj} $; \\
			$act \gets \Call{getPredicate}{obj} $; $tr = \{act, res, comp\}$; \\
			\If {$tr \notin ~kt$}{
				$kt \gets kt ~\cup~tr$; \\
			}
		}
	}
	
	\Return{$kt$};
\end{algorithm}

Algorithm \ref{algorithm:FindKws} describes how we extract keyword triplets. 
Specifically, each sentence is represented as a dependency tree after tokenization and lemmatization. We then retrieve the verbs and objects in the sentence based on the dependency information and POS tags of these words, which is similar to the process of getting \verb|SVOs|~\cite{findSVOs1,findSVOs2} in NLP. 
If these words are in our corpus, we will add them as well as their complements in the dependency tree into our keyword set. Take the aforementioned sentence ``Allow apps to send dump information'' as an example. 
First, function \emph{getVO} extracts phrases including ``allow apps'' 
and ``send information'' from the dependency tree of the sentence. But neither the verb ``allow'' nor the object ``apps'' is in our corpus via function \emph{InActionCorpus} at line 3 and \emph{InResourceCorpus} at line 11. So they will be dismissed by the parser. 
On one hand, the verb ``send'' is a common action in our corpus, thus we can get the resource ``information'' and its complement ``dump'' via function \emph{getObjs} at line 4 and \emph{getComp} at line 6 respectively. 
On the other hand, as ``information'' is in our resource corpus, function \emph{getComp} at line 11 and \emph{getPredicate} at line 12 return a keyword triplet: \{send, dump, information\}. 

Note that we do not extract the subject of one sentence since it is often omitted in a policy comment. However, we can determine the subject through the document filename. Last, all the comments in one \textsfs{*.te} file will be represented by two sets of keyword triplets: one from the comments of \emph{allow rules} and the other from those of \emph{neverallow rules}. At last, each keyword triplet is treated as a normalized sentence in the corresponding ``document'' (the \textsfs{*.te} file it locates). 
We use \textit{doc2vec}~\cite{mikolov2013distributed} to embed these comments into two 300-dimensional vectors for model training.

\subsection{Wide\&Deep Based Classification} 
\label{subsection:Wide Deep}

In this section, we describe how to augment our data to solve the data imbalance problem and how the models are trained.

\noindent\textbf{Data Augmentation. } 
After obtaining atomic rules from AOSP, we notice that the ratio of \emph{allow rules} and \emph{neverallow rules} is extremely imbalanced, which may bias the model.  For instance, 95.4\% of atomic rules in Android 8 are neverallow. 
That is because most of the \emph{allow rules} specify fine-grained domains, yet the \emph{neverallow rules} usually refer to a large number of domains so that only privileged domains can access sensitive data. 
To construct a training dataset with balanced labels, we manage to supply some additional atomic rules based on the negations in CIL files. 
Specifically, CIL introduces a set of attributes with the prefix of \emph{base\_typeattr}. These attributes are a subset of the attributes defined in AOSP.
For example, \emph{base\_typeattr\_293} is assigned to all the types with attribute \emph{appdomain} except type \emph{shell} and \emph{con\_monitor\_app} by the entry: ``\textit{typeattributeset base\_typeattr\_293 (and (appdomain) not (shell con\_monitor\_app))}''.
So from those \emph{neverallow} rules defined by \emph{base\_typeattr}, we can infer which subjects cannot perform such behavior except the subject excluded by \emph{base\_typeattr}. The \emph{base\_typeattr\_293} is not allowed to access files labeled by \emph{con\_monitor\_app}, which alludes that both \emph{shell} and \emph{con\_monitor\_app} can access such files.
Similarly, extra atomic allow rules could be inferred from negation statements in neverallow rules. These atomic rules inferred from negations can greatly enrich our training dataset.
Moreover, by limiting the number of ``inferred'' atomic rules, we can adjust the proportion of the labeled data in the training dataset.

\noindent\textbf{Redundancy Removal. }
Only the customized rules in the images will be considered for further classification. 
Therefore, we perform a comparison to obtain the customized part. 
Note that the representation of atomic rules can help us precisely find the manufacturers-defined rules regardless of the attributes' difference between devices. 
Last, 3.5 million atomic rules added by manufacturers are obtained from over 40 million atomic rules, which yields 267,162 unique atomic rules in total. 

\noindent\textbf{Wide\&Deep Model. } After obtaining the atomic rules, we jointly trained a wide linear model and a deep neural network to classify whether one rule is unregulated or not.
Considering the significant evolution of Android policy, 
we train one model for each Android version separately. 
The features extracted from the basic semantics in atomic rules and the User ID (\ie, F1 and F2) are encoded into a one-hot representation. They are fed for training a Logistic Regression (LR) classifier, the ``wide'' part of the model. 
In addition, some features may work better after a combination. For example, the \emph{class} field indicates the class of the \emph{object}, and it confines the \emph{permission} field as well. 
Crossing these fields together enables the wide model to treat them as a synthetic feature and learn the weight for each combination of them.

However, such a linear model is not precise enough because these features are discrete and extremely sparse - in SEAndroid policy, only a little amount of types are correlated. Most of the types are neither correlated by \emph{allow} nor \emph{neverallow} in the policy. 
It makes the traditional machine learning algorithms perform ineffectively on the classification of previously unseen subject-object pairs because they can only memorize the combinations of features that have appeared in the training set. 
Further, some rules may have indirect connectivity, resulting in that we might inevitably miss some inconspicuous but informative features in feature selection. 

To address these issues, we further represents the features into a low-dimensional dense embedding vector and then use a Deep Neural Network (DNN) to learn the relationship of unseen feature combinations.
The idea is inspired by the model adopted by modern recommender systems, in which a user is not correlated with most of the items but only shows his preference or dislike for a limited number of items in a large dataset.  
Recent studies~\cite{cheng2016wide,guo2017deepfm} solved this problem by jointly using a linear model (\eg, LR or Support Vector Machines) for learning the frequent co-occurrence of feature combinations in the training dataset, as well as a deep model (\eg, DNN) for exploring new feature combinations based on transitivity of correlation. This structure has proved its effectiveness in real-world Click-Through-Rate (CTR) prediction tasks such as recommending applications for users in Google Play. 
It predicts users' preferences based on user profiles. Similarly, we can also predict the subjects' ``preference'' to objects based on their features.

The DNN classifier usually takes the embedding representation of the data as input. 
We map the one-hot-encoded features F1 into a dense vector space via embedding and concatenate them with the embedding representation of policy comments features (F3). 
Then we use four fully connected layers to take these embeddings as input.
Finally, the deep model shares the same output unit with the liner model, which yields the final classification results of atomic rules. 
The details of the classification model and the input features are available on our website~\cite{modelcolumns}. 
When the training is finished, these models will receive the vectorized manufacturers atomic rules $r_a$ and output a classification result $\mathcal{DC}(r_a)$. The $r_a$ would be highlighted as unregulated ones if $\mathcal{DC}(r_a) \neq \phi(r_a)$.  

%% file: 5-evaluation.tex
\section{Evaluation}
\label{subsection:Extraction}
In this section, we evaluate the efficacy of \tool and compare it with the related approach in prior research.

\begin{table*}
  \centering
  \small
  \renewcommand\arraystretch{0.8}
  \caption{The performance metrics of \tool and \textsc{EASEAndroid}}\label{table:evaluation metrics}
  \vspace{-0.15in}
    \begin{tabular}{c|cc|ccc|cccc} \toprule
      \multirow{2} * {\textbf{Android Version}} & \multicolumn{2}{c|}{\textbf{Evaluation dataset}} & \multicolumn{3}{c|}{\textbf{SEPAL's metrics}} & \multicolumn{4}{c}{\textbf{Baselines metrics ($m,\sigma)=(10,55\%)$}}\\
      \cline{2-10}
      ~ & \# of Positive & \# of Negative & Accuracy & Precision & Recall  & Accuracy & Precision & Recall & \# of Unclassified\\
      \hline
      Android 5 & 19438 & 7375  & 0.986 & 0.990 & 0.991  & 0.895 & 0.966 & 0.893 & 742 (2.77\%) \\
      Android 6 & 7956 & 8893  & 0.990 & 0.969 & 0.973  & 0.790 & 0.633 & 0.747 & 1320 (7.83\%) \\
      Android 7 & 14438 & 27610 & 0.985 & 0.978 & 0.981  & 0.801 & 0.660 & 0.832 & 1488 (3.54\%)\\ 
      Android 8 & 118493 & 109670 & 0.984 & 0.986 & 0.983 & 0.831 & 0.934 & 0.868 & 1029 (0.45\%) \\ 
      Android 9 & 200107 & 152009  & 0.978 & 0.981 & 0.980 & 0.847 & 0.899 & 0.851 & 1432 (0.41\%)\\ \hline
      
      \textbf{Average} & 72086 & 61111 & 0.985 & 0.981 & 0.982 & 0.833 & 0.818 & 0.838 & 1202 (3.00\%) \\
      \bottomrule
    \end{tabular}%
\vspace{-0.2in}
\end{table*}

\subsection{Environment Setup}
\noindent\textbf{Implementation.} We build a prototype of \tool in Python, which contains 2 KLOC for atomic rule collection, 3 KLOC for feature extraction, and 500 LOC for the Wide\&Deep based classification. We use \textsc{spacy}~\cite{spacy}, a state-of-the-art NLP library for dependency parsing and Tensorflow Estimator~\cite{tensorflow} for model training.

\begin{figure}
  \centering
  \subfigure[Distribution of Android versions]{
  \includegraphics[width=0.21\textwidth]{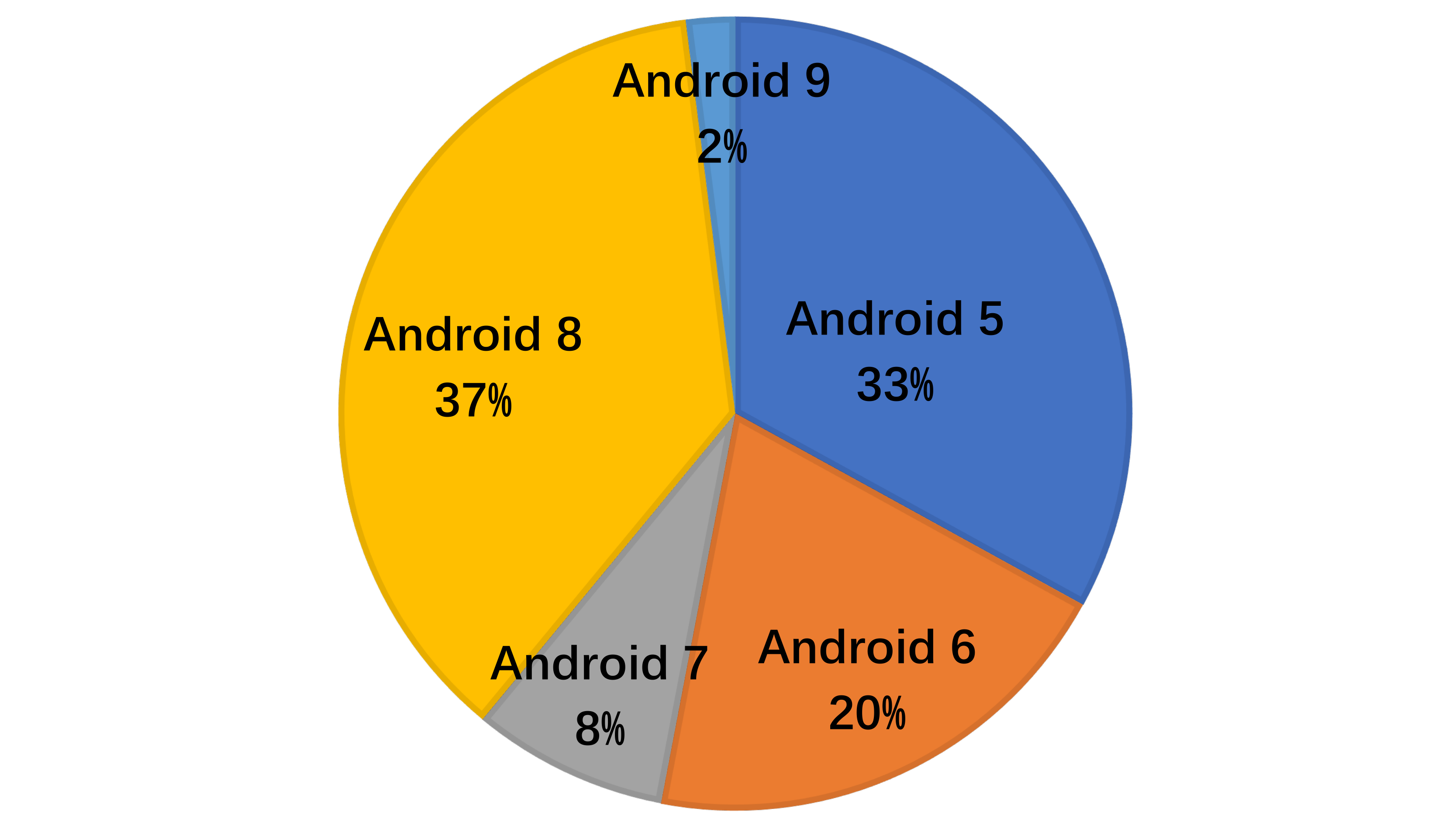} 
  }
\hfill
  \subfigure[Distribution of manufacturers]{
  \includegraphics[width=0.23\textwidth]{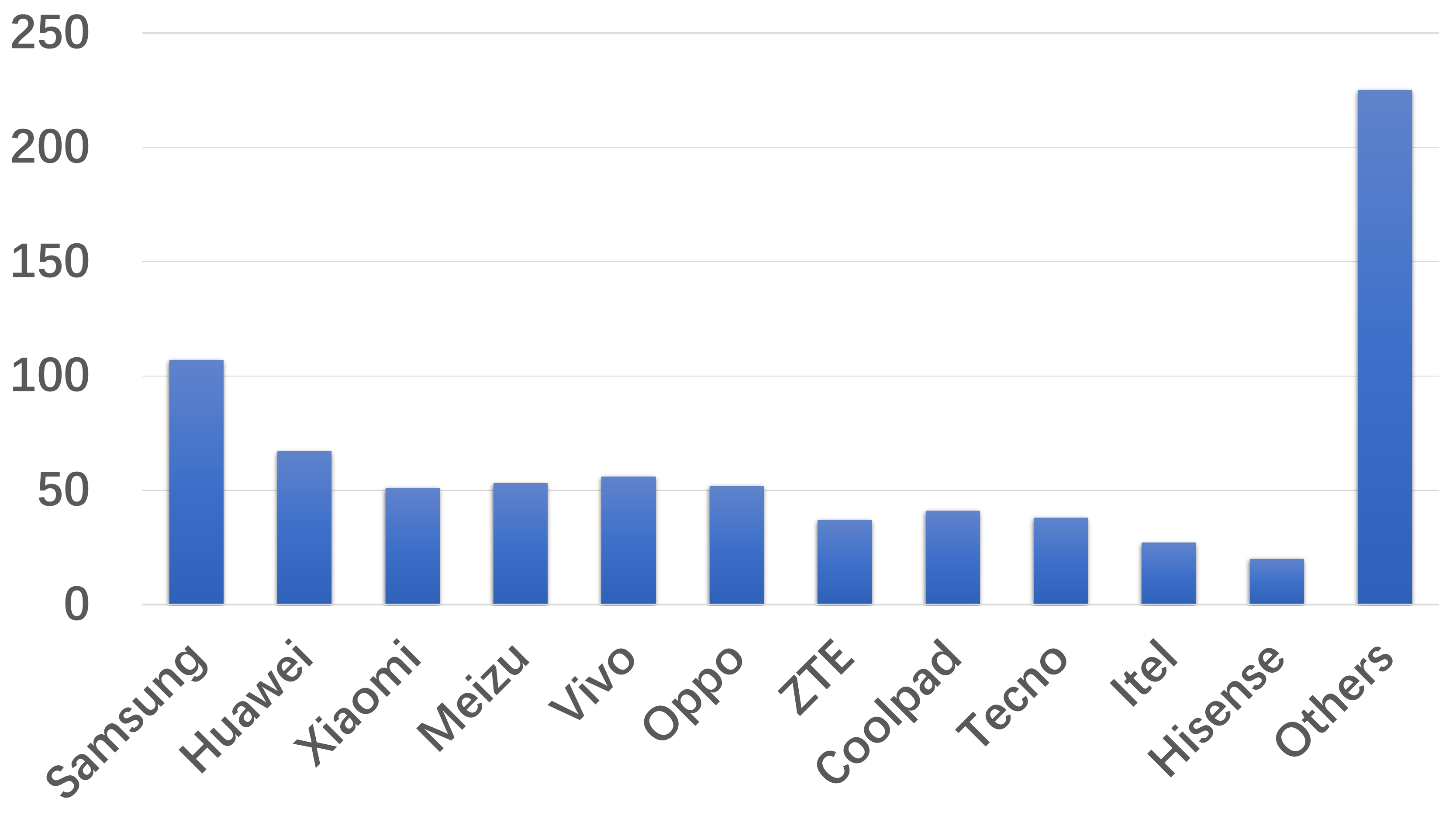}
  }
\vspace{-0.2in}
\caption{Distribution of firmware images}\label{fig:image distribution}
\vspace{-0.2in}
\end{figure}

\noindent\textbf{Dataset.}
Considering that the evolution of Android can significantly affect policy writing, we check out the latest released branch for each version (\ie, from Android 5 to Android 9) as the reference policy. 
In total, 27,668 original policy rules of \emph{allow} rules and 8,446 of \emph{neverallow} rules are obtained. 
After atomic rules extraction and data augmentation, 6.7 million of atomic rules are obtained. 
To gather customized rules for a large-scale study, we collect 774 stock Android firmware images from both manufacturers' official websites and the third-party forums like~\cite{xda-developers,AndroidMTK}. 
Figure~\ref{fig:image distribution} shows the distribution of collected firmware images for both AOSP and manufacturers. In total, over 3.5 million of atomic rules are obtained from 595,236 customized policy rules after redundancy removal. 

\subsection{Classification Accuracy}
\noindent\textbf{Comparison with baseline. } 
To illustrate the effectiveness of \tool,  for each Android version, we extract 10\% of the atomic rules as our test dataset. 
Then we use \tool to check whether the predicted result of a given atomic rule in the test dataset is consistent with its label, \ie, \emph{allow} or \emph{neverallow}. Additionally, we compare \tool with a baseline approach based on \textsc{EASEAndroid}~\cite{wang2015easeandroid}, which uses a Nearest-neighbors-based (NN) classifier and a Pattern-to-rule distance measure to refine the policy without NLP feature extraction and DNN classification. Specifically, it will first find all the ``neighbors'' of the target customized rules in the training dataset.  
The neighbors of one certain rule are determined if there is only one field difference between the rules. For example, two rules are neighbors if they are defined by different domains but share the same $\langle t, c, p \rangle$. 
Then the baseline algorithm classifies the target rule based on the label of the majority of neighbors in the training dataset. But if the number of the neighbors ($m$) is less than 10 or the majority of the neighbors occupies less than 55\% (\ie, $\sigma$ $<$ 55\%), it will classify the target as ``unclassified''.
Such thresholds are also adopted by \textsc{EASEAndroid}.
Note that if we use a stricter setting for the baseline approach, for example, increasing the values of $m$ and $\sigma$, it will introduce more unclassified rules. In our experiments, when we used the \emph{semi-auto mode} of \textsc{EASEAndroid}, \ie, threshold $(m,\sigma) = (10,75\%)$, there were 32\% of atomic rules remained unclassified in Android 6. The percentage would increase to 63\% if we use \emph{auto mode} ($(m,\sigma) = (10,85\%)$) defined by \textsc{EASEAndroid}.. 

Table~\ref{table:evaluation metrics} shows the results in terms of Android versions for \tool and the baseline method. 
\tool achieves 98.5\% accuracy, 98.1\% precision, and 98.2\% recall on average, and \textsc{EASEAndroid} (threshold $(m, \sigma) = (10,55\%)$) obtains an average of 83.3\% accuracy, 81.8\% precision, and 83.8\% recall upon the same test dataset. 
Thanks to the additional Deep Neural Network and the informative features, \tool outperforms the baseline significantly, not to mention that there are still some rules that the baseline cannot classify.

We manually analyze mistakes in the baseline results to recognize the gap between \tool and \textsc{EASEAndroid}.
On the one hand, the baseline algorithm blindly classifies the behaviors performed by unprivileged subjects into negative ones because these subjects appear quite frequently in neverallow rules. On the other hand, even the privileged subjects are restricted by some neverallow rules in AOSP, but the baseline algorithm classifies these forbidden behaviors as allow.
By a manual examination, we find it is difficult for \tool to classify behaviors that occur infrequently, such as the rules related to capability management (\eg, \emph{sys\_admin} and \emph{net\_admin}) and rarely used inter-process communication channels (\eg, shared memory \emph{shm} and message queue \emph{msg}). It is largely due to the lack of related training data, so that analysts have to verify these rules with rarely used classes and permissions manually.

\noindent\textbf{Manual Examination. }
To evaluate the practicality of \tool, we perform a comprehensive manual verification of the unregulated rules found in a physical device, \ie, Huawei P20 with Android 9.
We review the latest AOSP commit message, comments for these rules, the functionality and capability of related processes, and permission bits of associated resources, as well as the consultation of the policy developers.
Among 368 investigated atomic rules, 283 (76.9\%) of them are confirmed to be unnecessary and overly permissive. Nearly half of them in the test device are introduced due to the misuse of an attribute named ``hal\_audio'' (see Section~\ref{subsection:unregulated reasons}).
However, 
due to the lack of corresponding behaviors in the training dataset, \tool misclassified 40 (10.9\%) atomic rules into unregulated ones. 
For example, the rules allow \emph{kernel} to manage the Linux \emph{capabilities} including \emph{mknod}, \emph{sys\_admin}, and \emph{chown} are mistakenly classified as unregulated rules. 
The remaining 44 rules are difficult to determine without sufficient references and comments. 

%% file: 6-analysis.tex
\section{Result Analysis}
\label{section:A Large-scale Measurement}
In this section, we analyze the results presented by \tool and perform a large-scale measurement on the policy customization in the wild to answer the following three research questions (RQs).

\input{6-largescale_analysis}
\input{6-rq3}
\input{6-rq4}

%% file: 6-largescale_analysis.tex
\subsection{RQ1: How does the policy customization evolve across versions?}
\begin{figure*}[t]
	\centering
	\subfigure[The overall average number of (unregulated) rules and occupied percentages per Android versions]{
		\includegraphics[width=0.3\textwidth]{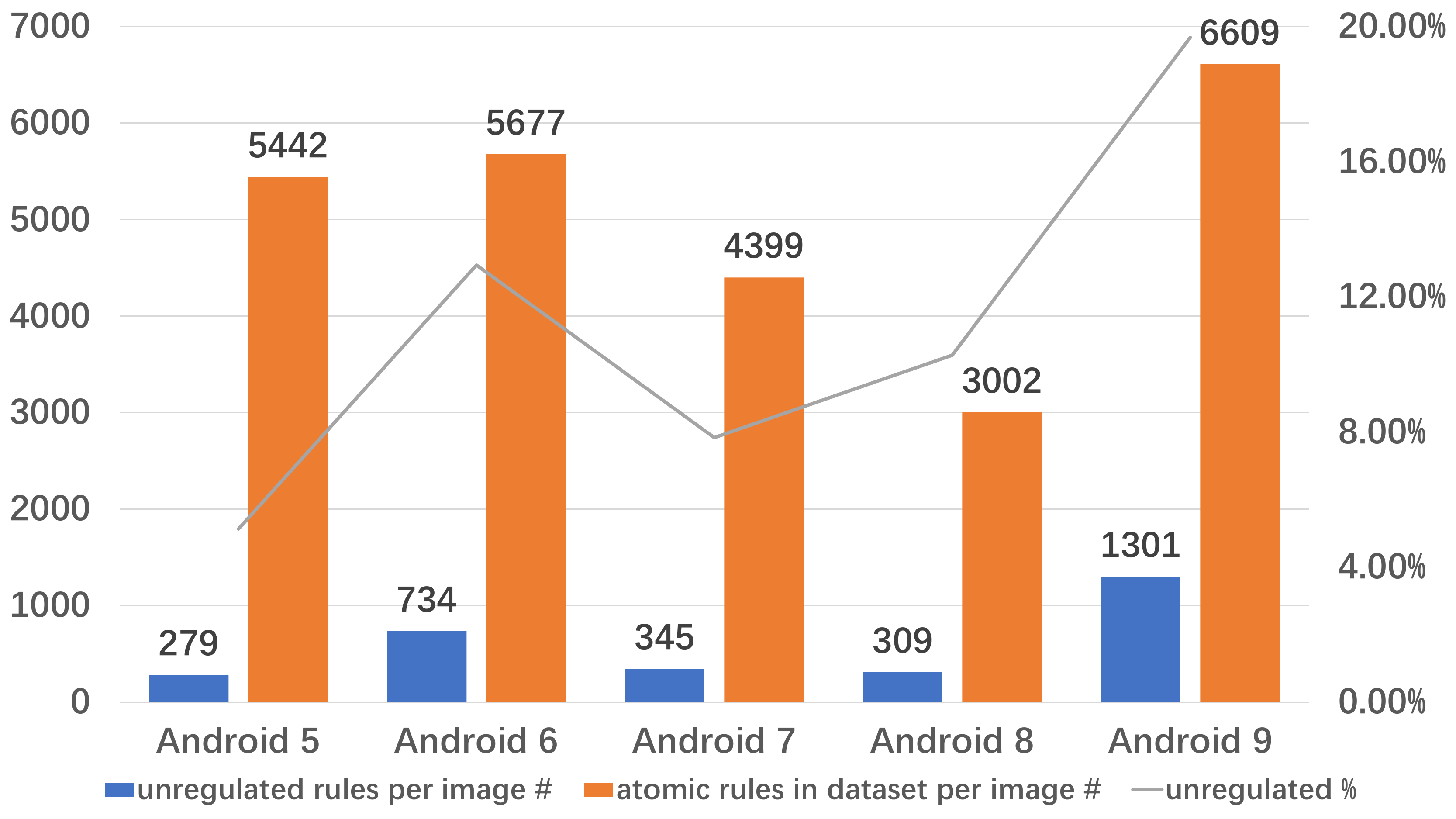} 
	}
	\hfill
	\subfigure[Distribution of Samsung's unregulated rules]{
		\includegraphics[width=0.3\textwidth]{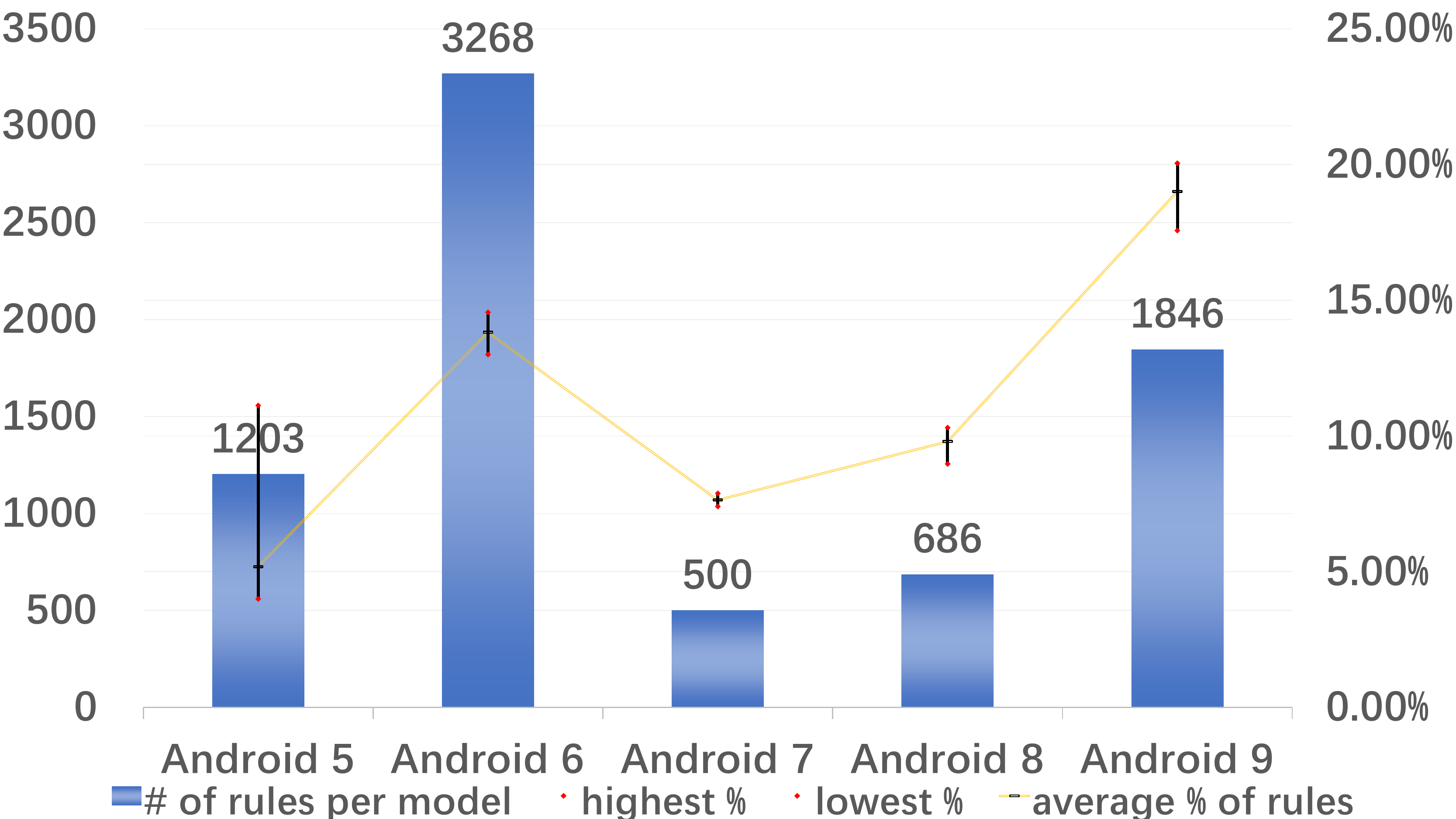} 
	}
	\hfill
	\subfigure[Distribution of Huawei's unregulated rules]{
		\includegraphics[width=0.3\textwidth]{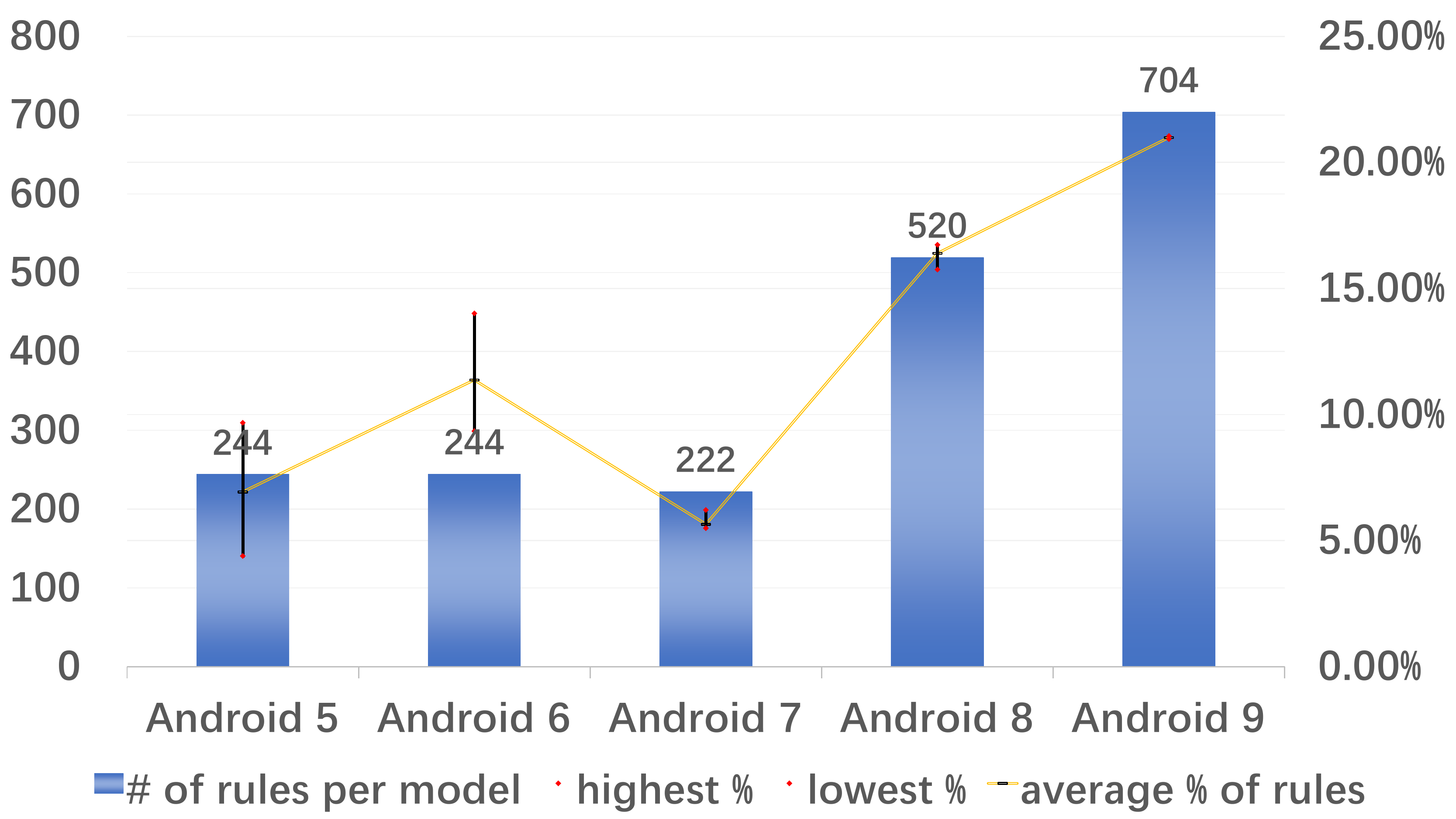}
	}
	\vspace{-0.2in}
	\caption{Distribution of the unregulated rules. The high-low lines indicate the percentage range}
	\vspace{-0.5cm}
	\label{fig:large-scale analysis}
\end{figure*}

\noindent\textbf{Overview. }
Fig~\ref{fig:large-scale analysis} shows how the number and percentage of unregulated rules evolve across the Android version and manufacturer.

In Android 5, the official policy is still lenient. For most manufacturers, they didn't have to perform heavy customization on AOSP rules.
On average, each image contains 279 (5.13\%) unregulated atomic rules.
In fact, across 13 manufacturers in the dataset of Android 5, Samsung and Huawei are the major contributors (94.57\%) to the unregulated rules in Android 5. For other manufacturers, less than one hundred atomic rules are added. 
Due to the lack of limitations on app processes asserted by \emph{neverallow rules}, 20.20\% of the rules in Android 5 are defined by app-related domains. Most of them are explicitly forbidden in the later versions of Android.
Besides that, a typical category of the unregulated rules in Android 5 is defined by coarse-grained attributes such as \emph{proc}, \emph{sysfs}, and \emph{unconfineddomain}. These rules respectively allow the subject to access nearly all the nodes in \textsfs{/proc}, \textsfs{/sys/fs} and any other mounted filesystems without a declared label. In the later versions of Android, Google defined many finer-grained types for the nodes in these filesystems and extremely limited access to proc and system partitions. 

In Android 6, though the average number of the customized atomic rules did not rise much (from 5442 to 5677),
the percentage of unregulated rules significantly increased from 5.13\% to 12.94\% in the 6.0 era - manufacturers such as ZTE, Meizu, and Oppo started to add more unregulated rules (nearly 20\%) at that time, most of which is defined by the coarse-grained attributes assigned for data partition and device nodes in \textsfs{/dev}.

In Android 7, some famous manufacturers broke down the coarse-grained attributes and types mentioned above. The reason might be that research had shown the security issues introduced in policy customization. For instance, problematic rules such as the overuse of \emph{unconfineddomain}, \emph{unlabeled}, and unprivileged app domains summarized by Reshetova \etal~\cite{reshetova2015characterizing} hardly appeared since Android 7. Thus the average number was reduced to 345 and the percentage was reduced to 7.83\%.

In Android 8 and Android 9, the official policy has become much finer than ever before. 
However, the percentage of unregulated rules has risen again. The percentage came to 10.27\% in Android 8 and even rise to 19.68\% in Android 9, which shows that the customization on the policy has not improved but become worse again. It has even become more and more difficult to keep up with the official policy for manufacturers. A possible reason is the introduction of Project Treble~\cite{projecttreble}, which separates the lower-level customized code from the Android system framework.
To update SEAndroid to work with Treble, manufacturers have to maintain their own hardware-specific policy rules and build their own images so that they can update those images independent of the AOSP, which is prone to introduce unregulated rules in devices. 
The vast bulk (27.61\%) of the unregulated rules defined for hardware abstraction layer (HAL) domains prove our concerns. These domains with the prefix \emph{hal} are assigned to the hardware layer processes implemented in vendor images since Android 8.

\noindent \textbf{Case study on Samsung. }
Samsung is the only manufacturer that customizes policy heavily in Android 5 devices. 
On average, each image contains 1,203 unregulated rules in Android 5 and the number even comes to 3,268 in Android 6, which is several times that of other devices.
Samsung introduced many coarse-grained rules at the early stage. For instance, nearly half of the rules are related to the use of socket. It could be caused by the overuse of common attributes, such as \emph{domain}, and \emph{unlabeled}, which leads to the introduction of many irrelevant atomic rules in one entry in the source code.
The situation has been improved since Android 7. The percentage of unregulated rules has been reduced to 7.89\% and the average number of unregulated atomic rules is only 500. 

Additionally, Samsung defines batches of rules with customized defined attributes. For example, \emph{newAttrs} is only spotted in Samsung devices. They play similar roles as \emph{base\_typeattrs} introduced in the official CIL policy in Android 8. These attributes are responsible for a large number of unregulated rules of Samsung devices in Android 5 and 6.
Most of them were removed in Android 7 by Samsung, but several attributes such as \emph{platformappdomain} are still retained in the latest versions. It shows that Samsung has made its efforts in refining the policy rules in the early days.

\noindent\textbf{Case study on Huawei. }
In Android 5, Huawei performs well on policy customization - only 4.3\% of the atomic rules are unregulated. 
Although Huawei does not perform as heavy policy customization as Samsung, the performance of Huawei is similar to that of Samsung in earlier Android versions. 
However, in Android 8, Huawei had little improvement but got difficult to keep up with the official policy evolution - 15.74\% of the atomic rules in Android 8 and 20.92\% of the atomic rules in Android 9 are classified as unregulated ones.
Among these rules, nearly 30\% of the unregulated rules introduced by Huawei in Android 8 are related to types that are prefixed with \emph{hal}. All these types introduced by Project Treble are designed to refine the hardware abstraction layer (HAL) in official Android 8.
Further, we notice that in some Huawei Android 8 devices, attribute \emph{domain} is allowed to read all the link files of the executables in the system via ``\textsfs{allow domain exec\_type lnk\_file read}''. The combination of two coarse-grained attributes yields hundreds of unregulated atomic rules at one time. 

\noindent\textbf{Miscellaneous. }
Further, to study the overall ecosystem of policy customization across manufacturers, we analyze the images from 72 unique manufacturers under the same Android version (Android 8). Figure ~\ref{fig:manufacturers top10} presents the top 10 manufacturers with the highest unregulated percentage in Android 8. 
MTN is the manufacturers with the highest unregulated percentage (18.08\%) in Android 8. Note that the average unregulated percentage is only 10.64\% in Android 8.
To the best of our knowledge, the devices produced by these top 10 manufacturers are popular among some developing countries. 
However, their amount of unregulated per image is only around two hundred, even lower than the average number (415) of Android 8.
Compared with the aforementioned notable manufacturers, these manufacturers make as little customization as possible but have an extremely high percentage of unregulated rules. We further noticed that most of these rules are identical with the unregulated rules spotted in earlier versions of famous manufacturers' devices. For instance, they still use coarse-grained attributes such as \emph{proc} and \emph{sysfs} in their hal-related rules. 

Furthermore, the high-low lines in Figure~\ref{fig:large-scale analysis} indicate the differences between the phone models produced by the same manufacturers. Note that the carrier may affect the policy of the same model.
For instance, in Android 5, the policy of Samsung Note 3 for XAS (USA), a.k.a. N900P, is different from that of MCT (Canada), a.k.a. N900W. The unregulated percentage of the MCT is five percents higher than that of XAS. Specifically, the significant differences in their policy are related to the app process. It could happen when writing corresponding rules for pre-installed applications.

%% file: 6-rq3.tex
\subsection{RQ2: Why are the unregulated rules introduced by policy developers? }
\label{subsection:unregulated reasons}
To figure out why these unregulated rules are introduced by policy writers and why they are classified as unregulated by \tool, we review their original policy rules, from which the atomic rules are extracted.
We identify 7,111 distinct original rules and then distill four main categories by manual analysis.
Note that due to the wide diversity of these rules forms, not all the unregulated ones could be attributed to these categories. 

\begin{figure}[tb!]
    \centering
    \includegraphics[width=0.78\linewidth]{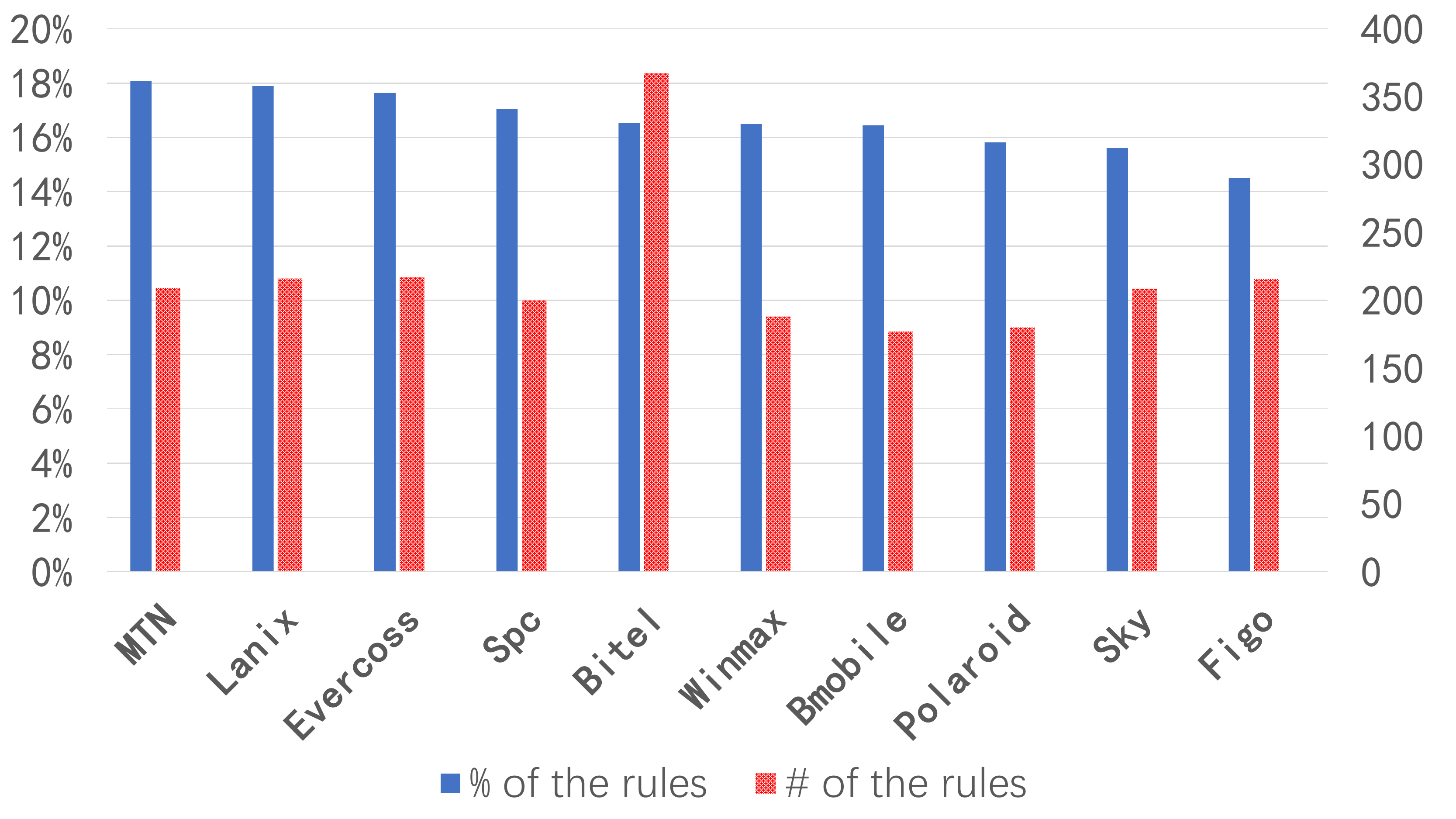}
    \caption{Top 10 manufacturers with the highest percentage}
    \vspace{-0.2in}
    \label{fig:manufacturers top10}
   
\end{figure}

\noindent\textbf{1. Misuse of attributes. }
After reviewing these original rules, we find that 692 of them are defined by coarse-grained attributes like \emph{appdomain} and \emph{netdomain}.
It suggests that many of the unregulated rules are likely to be introduced unintentionally by the policy writers. 
For example, a customized rule ``\textsfs{allow domain domain lnk\_file (read, getattr, open)}'' allows \textbf{ALL} processes in the system to open and read \textbf{ANY} link files in the \textsfs{/proc/[pid]/} directories.
Such a customized rule almost certainly yield over-privileged or unnecessary rules because tens of thousands of atomic rules will be introduced in one statement. This could happen when the policy writer simply accepted the output from \emph{audit2allow}, a tool that converts denials in the logs into corresponding policy statements. 
Another type of misuse happens in attribute assignment. 
For example, the AOSP rules below show that character device files labeled as \emph{audio\_device} are only allowed to be accessed by audio HAL domains. 
\begin{lstlisting}
# Declare attributes, assign them to domain "hal_audio_default" 
typeattribute hal_audio_default hal_audio
typeattribute hal_audio_default hal_audio_server
# Allow hal_audio_default to use audio device
allow hal_audio audio_device:chr_file rw_file_perms;
# Only audio HAL may access the audio hardware
neverallow { halserverdomain -hal_audio_server} audio_device:chr_file *;
\end{lstlisting}
However, in some Huawei devices, policy writer assigns the attribute \emph{hal\_audio} to domain \emph{hal\_ir\_default} via \textsfs{typeattribute hal\_ir\_default hal\_audio}'', which enables \emph{hal\_ir\_default} to gain a prohibitive privilege from the AOSP rules above. To the best of our knowledge, domain \emph{hal\_ir\_default} is used by an infrared-related server process. Since the corresponding code is not available, we cannot figure out why the infrared need to access the audio device. Even if it does need it, \emph{hal\_ir\_default} should only be granted the permissions it needs, instead of granting all the permissions hold by \emph{hal\_audio}.

\noindent\textbf{2. Testing and debugging related rules. }
Some of the rules in AOSP are created for the purpose of factory testing or debugging. These rules are usually declared by some macros such as \emph{build\_test\_only} and \emph{userdebug\_or\_eng}, which are only compiled in \emph{userdebug} or \emph{eng} builds. 
However, we noted that 308 manufacturer-defined original rules recognized as unregulated ones are actually defined by these macros in AOSP, which should not appear in release versions. Table~\ref{table:examples of debug rules} lists some examples of the rules.

\begin{table*}
  \caption{Examples of testing and debugging related rules}
  \vspace{-0.2in}
  \begin{center}
    \newcommand{\tabincell}[2]{\begin{tabular}{@{}#1@{}}#2\end{tabular}}
    \resizebox{\textwidth}{!}{
      \begin{tabular}{lll} 
        \toprule
        \textbf{Examples of unregulated manufacturers rules} & \textbf{Corresponding definition in AOSP} & \textbf{Annotations}\\
        \midrule
        allow su * & All the \emph{su-related} definition are defined by \emph{userdebug\_or\_eng}. & \emph{su} is assigned to superuser process, which should not appear in released version. \\
        allow audioserver media\_data\_file * & userdebug\_or\_eng(`allow audioserver media\_data\_file:dir create\_dir\_perms;') &  access debug related files for TEE sink - pcm capture. \\
        allow untrusted\_app perfprofd\_data\_file file * & userdebug\_or\_eng(`allow untrusted\_app\_all perfprofd\_data\_file:file r\_file\_perms;') & access perfprofd output in /data/misc/perfprofd/. \\
        allow mediaserver mediaserver process ptrace & userdebug\_or\_eng(`allow mediaserver self:process ptrace;')& ptrace to itself for memory leak detection. \\
        allow *\_app heapdump\_data\_file file *&userdebug\_or\_eng(`allow appdomain heapdump\_data\_file:file append;') & dumpsys related functions about sending heap dumps to system\_server \\
        \bottomrule
      \end{tabular}
    }
    \label{table:examples of debug rules}        
  \end{center}
  \vspace{-0.2in}
\end{table*}


\noindent\textbf{3. Deprecated rules. }
As we illustrated in Section~\ref{subsection:policy diversity}, the policy may change significantly along with the evolution of the Android system. 
Rules that are incompatible or possibly risky are likely to be removed from AOSP. Retention of these rules may pose risks for Android devices.
In our dataset, we find 393 occurrences of deprecated rules in the versions where they should not be present, 193 of which for Android 6, 50 for Android 7, 104 for Android 8, and 46 for Android 9.  
It suggests that manufacturers may not update their SEAndroid policy in time.   
For instance, rule ``\textsfs{allow init kernel:security load\_policy}'' allows \emph{init} to reload sepolicy in Android 4.3. This rule is utilized to bypass SEAndroid in the wild~\cite{ksma}, thus has already been removed from Android 6. However, it is still spotted in some devices of Android 7 and Android 8. 
It is known that the Android device can patch the known vulnerabilities through the monthly device updates, however, it seems not easy for manufacturers to update their SEAndroid policy rules in time. 

\noindent\textbf{4. Excessive permissions to untrusted domains. }
Domains such as \emph{untrusted\_app} and \emph{isolated\_app} are assigned to third-party app processes. Theoretically, these apps should not require any device-specific rules. However, nearly 10\% (3,287) of the unregulated atomic rules are related to these domains like the below rules. 
\begin{lstlisting}
#Read runtime information in procfs:
allow untrusted_app proc_stat file (read)
allow untrusted_app kernel file (read)
# Read runtime information in sysfs:
allow untrusted_app sysfs_net file (read)
# Access to device node:
allow untrusted_app input_device chr_file (ioctl)
# Access to system service:
allow untrusted_app meminfo_service service_manager (find)
\end{lstlisting}
Note that prior research~\cite{Chen:2017:ASP:3134600.3134638} detects privileged file access issues via a hand-crafted label set of privileged ordinary files. 
However, the boundary of ``privileged'' and ``unprivileged'' is blurred, especially when the number of types has significantly increased with the evolution of SEAndroid, while our learning-based method helps us get rid of the difficulty of manually constructing privileged file set.

%% file: 6-rq4.tex
\subsection{RQ3: What are the security impacts of unregulated rules? }
\label{subsection:RQ4}
The official documents of SEAndroid~\cite{writepolicy} claim that unnecessary rules can lead to the waste of memory and disk space, and also prolong the runtime policy lookup times. 
Besides performance, we are more concerned about the potential security impacts caused by these unregulated rules.
According to the documents~\cite{writepolicy}, SEAndroid takes effects in the following ways: 1) protect and confine system services; 2) control access to application data and system logs; 3) reduce the effects of malicious software; 4) protect users from potential flaws in code. However, all of them can be compromised by the unregulated rules we found. 
In this section, we demonstrate how the unregulated rules downgrade these original defenses mentioned above accordingly. In addition, we implement two attacks as proof of concepts, one of which is acknowledged by the manufacturer and patched in the later versions. 

\noindent\textbf{1. System service exposure. }
Services are designed to supply cross-application functionality in Android. Permissions ``add'' and ``find'' are defined upon class \emph{service\_manager}
to confine service registration and handler acquisition. A robust policy should well protect and confine system services.
However, some privileged services are found to be accessed by user-level processes. 
We find that system services such as \emph{meminfo service}, and \emph{lock\_settings\_service} are exposed to \emph{untrusted\_app} in some Android 8 devices, granting extra permissions to third-party apps. 
Here, we use the lock setting service as an example to show how the rule leads to an attack. 

\noindent\underline{Locksetting Attack. }
The ``\emph{lock\_setting\_service}'' is a type assigned to the screen lock pattern/password related processes. By communicating with the service, one client can set the lock pattern/password or test the existing password set by the user. Such a sensitive service should only be available for trusted processes. But an unregulated rule ``\textsfs{allow untrusted\_app lock\_settings\_service (service\_manager (find))}'' is spotted in an ASUS device of Android 8.
To evaluate the security impact of this rule, we prepared an Android (AArch64) emulator for testing. We added this rule into the emulator and supposed that the attacker had full control over an untrusted app process, which could be achieved by installing a malicious application on the target device. 
To communicate with the target service, we utilized the ``\emph{locksettings}'' program provided by stock Android. 
However, the service process contains a UID-based check that only allows root (uid=0) or shell (uid=2000) users to use the lock setting service. To bypass the DAC checks in the service, we develop an exploit using CVE-2017-7533 to modify the user id to 0. 
After that, we can successfully set and reset the pattern/password in our app process. 

\noindent\textbf{2. Sensitive file exposure. }
An unregulated rule may compromise the protection of sensitive files provided by SEAndroid.
Files located in ``\textsfs{sysfs}'' and ``\textsfs{procfs}'' partitions reflect the system status.
Reading these files may not lead to direct damage, but can be leveraged to perform a side-channel attack of inferring text input, apps running status in the foreground~\cite{simon2016don,diao2016no}, and behaviors performed in a browser~\cite{jana2012memento}. 
Rules such as reading the information of process \emph{zygote} may leak the memory layout and the runtime status of system daemons, which is a necessary step for vulnerability exploitation.
Furthermore, we notice that in some Samsung and Panasonic devices, third-party apps are allowed to run \emph{dumpstate}, a tool designed for root/shell users to monitor the status of all the processes. It could even be used to take screenshots without notification. 

\noindent\underline{Camera Driver Attack. }
Device nodes in \textsfs{/dev} (\eg, \emph{video\_device}, \emph{input\_device}) are the user interfaces to the hardware-specific device drivers. 
Processes in the application layer should only be allowed to communicate with these privileged drivers through framework services. However, we find these character files are exposed to \emph{untrusted\_app} in some devices:
``\textsfs{allow untrusted\_app video\_device chr\_file (ioctl read write getattr append open)}''.
In SEAndroid, type \emph{video\_device} is assigned to character files of the camera device node in the \textsfs{/dev} directory of the system. In theory, exposing these files to applications allows app processes to use the camera without any permission. However, it is non-trivial in practice because the hardware code is not open source. Without knowing how the driver works, we cannot control the camera. But a denial-of-service attack could be launched based on the ioctl commands defined in the V4L2 framework. The invalid parameters sent to the camera driver via \emph{ioctl} will crash and reboot the device. In our dataset, such rules are spotted in 26 images, of which 25 are found in Samsung Android 5 devices and the other one is found in a Kaicom Android 8 device. 

Prior research~\cite{zhou2014peril} shows that in some earlier Samsung devices, these character files are not protected by Linux DAC (the permission bits are 666). 
It means unregulated rules will completely expose these character devices to attackers. Even if the permission bits of DAC is configured correctly, the unregulated rules can also lead to an attack when the DAC is compromised.

\noindent\textbf{3. Capability extension of malicious apps. }
Object classes such as \emph{capability} and \emph{system} can be used to manage the Linux capabilities granted to root processes. These classes can provide powerful capabilities when the subjects are compromised~\cite{zygote2init}. However, we find that some unregulated rules even grant these capabilities to third-party apps.
For instance, they allow the untrusted domains to load kernel modules, make stack/heap of \emph{init} executable, and control system network via \emph{net\_admin} capability as follows:
``\textsfs{allow untrusted\_app untrusted\_app capability (net\_admin setuid setgid)}'',
``\textsfs{allow isolated\_app init process (execstack execheap rlimitinh sigkill setsched)}''.
Such behaviors have nothing to do with an ordinary app but can provide a great convenience for malicious apps to control the system. All these rules are spotted in Redmi 3s and ZTE BA520 obtained from the same website. 
We inferred that these ``stock'' images might be repacked and uploaded by malicious ROM developers thus we reported our concerns to the website manager. 

\noindent\textbf{4. Feasible paths for vulnerability exploitation. }
One of the key principles of SEAndroid is providing careful attack surface management to prevent vulnerabilities from being exploited. However, customized rules may enlarge attack surfaces. 
There is a variety of socket families in the Linux kernel, and an abuse of these sockets will offer an additional opportunity for attackers to escalate privilege~\cite{shao2016the}. 
For example, CVE-2017-7184 is a vulnerability in Linux kernel that requires the use of \emph{netlink\_xfrm\_socket} by process \emph{netmgrd}. It means attackers cannot exploit this vulnerability unless they compromise process \emph{netmgrd}. However, we find that some app domains are allowed to use this socket in manufacturers' devices, which makes it possible to trigger the issue from unprivileged processes.
Rules related to other families such as netlink route sockets, ping sockets and unix stream sockets are also found in some earlier Samsung devices. 
These rules can be utilized as potential escalation paths in the kernel vulnerability exploitation once the corresponding domains are compromised.

%% file: 6-discussion.tex
\section{Discussion}
\label{discussion}

\noindent\textbf{Response from vendors. }
We have reported unregulated rules found in the devices of the latest Android versions (\ie, Android 8 and 9) to seven vendors--ASUS, Haier, Huawei, Meizu, Oppo, Samsung, and Xiaomi. As of this submission, we have received replies from four vendors, all of which confirm our findings. 
Some unregulated rules have been removed from their latest devices as suggested, and the others are retained. 
According to the feedback from these vendors, we summarize two difficulties for vendors in handling these unregulated rules: 
1) \emph{Unknown security impacts by these rules}. Due to complexity, it is challenging to measure the influence of one single rule on the system. This can be mitigated if a proof-of-concept is developed for validation. 
2) \emph{Unpredictable dysfunction by rule removal}. Since one rule is used to regulate the permissive behaviors of subjects, removing them may inadvertently cause system failure or degradation. Therefore, developers have to adjust the affected subjects if one rule is removed.
However, sometimes even the system developers may not understand why these rules are introduced.
Given this, it is urgently desired to identify the security hazards of insecure rules and patch policy automatically, which is a promising research direction.

\noindent\textbf{Significance.}
Our study, together with the tool, is beneficial for three stakeholders.
Manufacturer policy developers can use \tool to quickly identify the risky or unnecessary rules introduced during customization. 
AOSP policy developers can refer to the issues confirmed by manufacturers and make new neverallow rules to enhance security. 
Security analysts can use \tool to identify the extra attack surfaces introduced in Android customization.
For example, the camera driver attack in Section~\ref{subsection:RQ4} shows that unregulated rules can triage the misconfigurations in DAC. 

\noindent \textbf{Limitations. } 
Our approach suffers from some limitations. In particular, \tool relies on official policy rules for training. An error in these rules, even with a low probability, can negatively influence our training model.
Moreover, our approach cannot cope with newly-defined types. This is because \tool is a learning-based approach, and inherently susceptible to the data that is never seen before. To mitigate this, we can enhance the learned model by training more rules introduced by both AOSP and OEMs. 
Third, we can only extract the DAC information for subjects from configuration files ``\textit{*.rc}'' without a physical device. It can degrade our model training. 
Even though a recent work \textsc{BigMAC}~\cite{grantbigmac} can recover an approximate filesystem state of a running system based on a firmware image, DAC checks exist in the non-file objects are mot explicitly declared thus are hard to identify (such as the DAC checks in the lock setting service). 
Further, from the perspective of attacker, unregulated rules may not lead to real world attacks directly because other security mechanisms such as DAC and seccomp can hinder the process of exploitation. But from the perspective of system developers, SEAndroid should hold even when the DAC fails~\cite{smalley2013security}. 

%% file: 7-relatedwork.tex
\section{Related Work}
\noindent\textbf{Study on SELinux. }
A line of research~\cite{jaeger2003analyzing,sarna2004policy,hicks2010logical,sasturkar2006policy} performs security analysis on SELinux to figure out the unintended policy rules.
Zanin \etal~\cite{zanin2004towards} use a formalization way to represent SELinux policy model and detect the conflicts in the rules. 
Guttman \etal~\cite{guttman2005verifying} and Jaeger \etal~\cite{jaeger2006prima} further perform information flow analysis to measure the policy. 
These studies try to harvest the conflicts in rules via a predefined trusted base while we aim to find the unregulated type pairs that should not be associated from a statistical perspective.
Moreover, SEAndroid has an entirely different architecture and attack surface, making traditional methodology for SELinux not suitable for Android. Performing system-level information flow analysis is also not realistic for commercial Android devices since we cannot modify the system as we need. 

\noindent\textbf{Study on SEAndroid. }
Reshetova~\etal~\cite{reshetova2015characterizing} and Chen \etal~\cite{Chen:2017:ASP:3134600.3134638} manually analyze SEAndroid policy rules from several devices and propose some problematic patterns of the unregulated rules. 
Analyzing the policy rules based on the predefined patterns can get highly explainable results, but it is extremely dependent on expertise, and it cannot actively discover issues unknown to experts. 
\textsc{BigMAC}~\cite{grantbigmac} is a recent study that combines all layers of the DAC/MAC policy together in a graph. It performs analysis on firmware images and can recreate the security state of a running system. 
However, analysts might be confused about what to query, 
but our machine learning method can automatically highlight the most suspicious ones and then \textsc{BigMAC} can be used for further manual analysis. 
Im \etal~\cite{im2018historical} performs a historical study on the evolution of the AOSP policy. They call for the new technology to analyze a large number of complex rules, which is exactly the same for our study.
Some of the related researches focus on the refinement of policy via runtime logs and tests. \textsc{EASEAndroid}~\cite{wang2015easeandroid} uses machine learning to analyze million-level audit logs and generate new rules automatically. The primary purpose of their research is different from ours. They mainly focus on refining existing policy but we aim to find new issues of the customized policy in the wild. 
Wang \etal~\cite{wang2017spoke} performs a knowledge collection on SEAndroid based on a function test toward Android application and framework. They use the knowledge base for characterizing the attack surface of policy.

%% file: 8-conclution.tex
\section{Conclusion}
In this paper, we propose \tool to help manufacturers examine their customized rules effectively. \tool learns the relationships among official types by utilizing deep learning and NLP techniques. 
The evaluation shows \tool can help to highlight the rules that deserve attention among the massive number of customized rules. It also yields several prior known issues that are recognized by multiple vendors. 
Even though, SEAndroid has been deployed for years, the large-scale measurement results suggest that the policy writing in the wild is still in a mess. 
We hope our findings could help to improve the customization of SEAndroid policy.

%% file: Acknowledgement.tex
\section*{Acknowledgement}
We thank the anonymous reviewers for their comprehensive feedbacks. This research is supported in part by Key Laboratory of Network Assessment Technology, Chinese Academy of Sciences, Beijing Key Laboratory of Network security and Protection Technology, Strategic Priority Research Program of CAS (No.XDC02040100), NSFC (No.61802404, 61802394, 61902396, 61902395, and U1836211), Beijing Natural Science Foundation (No.JQ18011), Youth Innovation Promotion Association CAS, and CCF-Tencent Open Fund.